\PassOptionsToPackage{table,xcdraw,dvipsnames}{xcolor}
\documentclass[sigconf,screen]{acmart}
\usepackage{tcolorbox}
\usepackage{listings}
\usepackage[flushleft]{threeparttable}
\usepackage{multirow}
\usepackage{enumitem} 
\usepackage{setspace}
\usepackage[skip=1ex]{caption}
\newcommand{\tool}{\textsc{VulTrial}}

\definecolor{myblue}{RGB}{230, 240, 255} 
\definecolor{myborder}{RGB}{100, 100, 255} 
\usepackage[english]{babel}

\AtBeginDocument{%
  }

\copyrightyear{2026}
\acmYear{2026}
\setcopyright{cc}
\setcctype{by}
\acmConference[ICSE '26]{2026 IEEE/ACM 48th International Conference on Software Engineering}{April 12--18, 2026}{Rio de Janeiro, Brazil}
\acmBooktitle{2026 IEEE/ACM 48th International Conference on Software Engineering (ICSE '26), April 12--18, 2026, Rio de Janeiro, Brazil}
\acmPrice{}
\acmDOI{10.1145/3744916.3773256}
\acmISBN{979-8-4007-2025-3/2026/04}

\begin{document}

\title{Let the Trial Begin: A Mock-Court Approach to Vulnerability Detection using LLM-Based Agents}

\author{Ratnadira Widyasari\textsuperscript{$\diamondsuit$}, Martin Weyssow\textsuperscript{$\diamondsuit$}, Ivana Clairine Irsan\textsuperscript{$\diamondsuit$}, Han Wei Ang\textsuperscript{$\spadesuit$}, Frank Liauw\textsuperscript{$\spadesuit$} \\Eng Lieh Ouh\textsuperscript{$\diamondsuit$}, Lwin Khin Shar \textsuperscript{$\diamondsuit$}, Hong Jin Kang\textsuperscript{$\clubsuit$}, and David Lo\textsuperscript{$\diamondsuit$}}

\affiliation{%
  \institution{\textsuperscript{$\diamondsuit$}School of Computing and Information Systems, Singapore Management University, Singapore}\country{}
}
\affiliation{%
  \institution{\textsuperscript{$\spadesuit$}GovTech, Singapore}\country{}
}
\affiliation{%
  \institution{\textsuperscript{$\clubsuit$}School of Computer Science, University of Sydney, Australia}\country{}
}

\affiliation{%
  \institution{\{ratnadiraw.2020, mweyssow, ivanairsan, elouh, lkshar, davidlo\}@smu.edu.sg, \\ 
  \{ang\_han\_wei, frank\_liauw\}@tech.gov.sg, hongjin.kang@sydney.edu.au}\country{}
}

\renewcommand{\shortauthors}{Widyasari et al.}

\begin{abstract}
Detecting vulnerabilities in source code remains a critical yet challenging task, especially when benign and vulnerable functions share significant similarities. In this work, we introduce \tool{}, a courtroom-inspired multi-agent framework designed to identify vulnerable code and to provide explanations. It employs four role-specific agents, which are \emph{security researcher}, \emph{code author}, \emph{moderator}, and \emph{review board}. Using GPT-4o as the base LLM, \tool{} almost doubles the efficacy of prior best-performing baselines. Additionally, we show that role-specific instruction tuning with small quantities of data significantly further boosts \tool{}'s efficacy. 
Our extensive experiments demonstrate the efficacy of \tool{} across different LLMs, including an open-source, in-house-deployable model (LLaMA-3.1-8B), as well as the high quality of its generated explanations and its ability to uncover multiple confirmed zero-day vulnerabilities in the wild.
\end{abstract}



\maketitle
\section{Introduction}
Vulnerability detection in software code is a critical task to ensure system security and prevent potential exploits~\cite{chen2023diversevul}. Manual vulnerability detection can be time-consuming and error-prone, especially given the scale and complexity of modern software systems. 
To automate this, researchers have explored various machine learning and deep learning approaches~\cite{li2018vuldeepecker, russell2018automated, chernis2018machine, risse2024uncovering, zou2021interpreting, chakraborty2021deep}. 
Recently, with substantial advancements in large language models (LLMs) and their impressive performance on natural language tasks~\cite{wang2023can, abbasiantaeb2024let, zhuang2023toolqa, neumann2024llm}, several studies have found promising outcomes using LLMs for automated vulnerability detection~\cite{tamberg2024harnessing, ullah2023llms, zhou2024large_emerging, nong2024chain, zhang2024prompt, widyasari2024beyond}

Despite these advances, recent research by Ding et al.~\cite{ding2024vulnerability} highlighted significant limitations when evaluating current LLM-based vulnerability detection approaches. Their study proposed a challenging new benchmark called PrimeVul, where the functional differences between benign and vulnerable code pairs are subtle, less than 20\%. These cases where the code has a high degree of similarity are critical to evaluate because in real-world scenarios,  vulnerabilities often stem from subtle, non-obvious code changes~\cite{ni2023distinguishing,tan2024similar}. On this dataset, LLM-based approaches exhibited low performance. The best performance comes from using GPT-4~\cite{openai2023gpt4}, which achieves a Pair-wise Correct Prediction (P-C) score of 40 (out of a possible 435). P-C measures how many vulnerable-benign code pairs a method classifies entirely correctly, directly reflecting its ability to detect the subtle code changes that introduce a vulnerability. 
This LLM-based approach used in PrimeVul was implemented under the conventional single-agent system, wherein one model attempts to identify and explain vulnerabilities on its own. 
Single-agent systems are inherently constrained by their reliance on a single perspective when analyzing vulnerabilities, which can lead to issues such as hallucination and overconfidence in incorrect reasoning~\cite{zhang2023siren, shi2025mitigating, huang2025survey, liang2023encouraging, du2023improving}.  
These single-agent systems may fail to critically verify their own output, which is crucial when identifying subtle vulnerabilities~\cite{liang2023encouraging}.

Given these limitations, recent research has begun exploring multi-agent systems to automate software engineering tasks~\cite{he2025llm, hu2023large, wei2024llm}. 
Such approaches distribute tasks across role-specific agents while allowing discussion and richer information exchange between agents.  
In this work, we propose \tool{}, a novel multi-agent LLM framework that employs a courtroom-inspired scenario. \tool{} features four role-specific agents: the \emph{security researcher} (prosecutor), \emph{code author} (defense attorney), \emph{moderator} (judge), and \emph{review board} (jury). 
The \emph{review board} bases its decision on arguments from the \emph{security researcher} (highlighting potential vulnerabilities), the \emph{code author} (defending the code), and the \emph{moderator} (summarizing debate points). 
The \emph{review board agent} works similarly to a jury in the courtroom scenario, where they need to hear all the evidence presented in the case and make a final verdict based on it. 
In \tool{}, we use the same underlying LLM for all agents, building on prior works~\cite{wei2022chain,chatterjee2024posix,hu2023large} that show prompt-induced roles can lead to different reasoning paths using the same model.

We benchmarked \tool{} against single-agent and multi-agent baselines. 
For the single-agent, we include deep learning (DL)-based models (CodeBERT~\cite{feng2020codebert}, CodeT5~\cite{wang2021codet5}, UniXCoder~\cite{guo2022unixcoder}, and LineVul~\cite{fu2022linevul}), as well as an LLM-based method (Ding et al.'s CoT prompts~\cite{ding2024vulnerability}). For the multi-agent setting, we utilized GPTLens~\cite{gptlens}. All LLM-based methods are instantiated using GPT-3.5 and GPT-4o as the base model (see Section~\ref{sec:model}).
The results show that \tool{} improves over the baselines. With GPT-4o, \tool{} nearly doubles the performance of both the best performing single-agent and multi-agent baselines (P-C score=81 vs. 40 and 44, respectively).

To assess the effect of instruction tuning on role-specific agents, we tuned each agent in \tool{} using 50 validation sample pairs. Instruction tuning the \emph{moderator agent} in \tool{} improves the P-C score by 15 (18.52\%) over the non-tuned version and surpasses the single-agent and multi-agent baselines by 56 (140\%) and 52 (118\%).
To assess generalizability, we additionally confirmed the efficacy of \tool{} using LLaMA-3.1-8B, a publicly available model that can be deployed in-house. 
Separately, we perform an ablation study to verify that each agent contributes positively to achieving the best performance. Beyond quantitative results, we also qualitatively evaluate the generated explanations, examining completeness, clarity, actionability, and informativeness, and observe strong performance across all four dimensions. Finally, we deploy \tool{} in the wild, where it successfully uncovers multiple confirmed zero-day vulnerabilities, demonstrating its practical effectiveness.

To summarize, the contributions of our study are as follows:
\begin{itemize}[leftmargin=0.05\linewidth, itemsep=1pt, parsep=0pt, topsep=1pt, partopsep=0pt]
    \item We propose \textbf{\tool{}}, a novel multi-agent, courtroom-inspired approach leveraging LLMs for automated vulnerability detection. Also, different from many prior approaches that only output a binary label, 
    \tool{} provides {\em detailed explanations} for each classification. 
\item We demonstrate that targeted instruction tuning of role-specific agents in \tool{} improves the performance.
\item We conduct extensive experiments demonstrating the efficacy of \tool{} on benchmark data and in the wild. 
\end{itemize}

\section{\tool{}}\label{sec:approach}
\tool{} leverages multiple LLM-based agents modeled after courtroom roles: a \emph{security researcher} (prosecutor), \emph{code author} (defense attorney), \emph{moderator} (judge), and \emph{review board} (jury). This structured debate enables thorough, adversarial discussions for deeper analyses and refined conclusions regarding code vulnerabilities. 
Drawing inspiration from courtroom practices, where structured adversarial processes have been shown to reduce wrongful convictions and improve the fairness of judgments~\cite{findley2006multiple, garrett2011convicting}, \tool{} aims to improve vulnerability detection in a similar manner.

\begin{figure}
    \centering
    \includegraphics[width=0.78\linewidth]{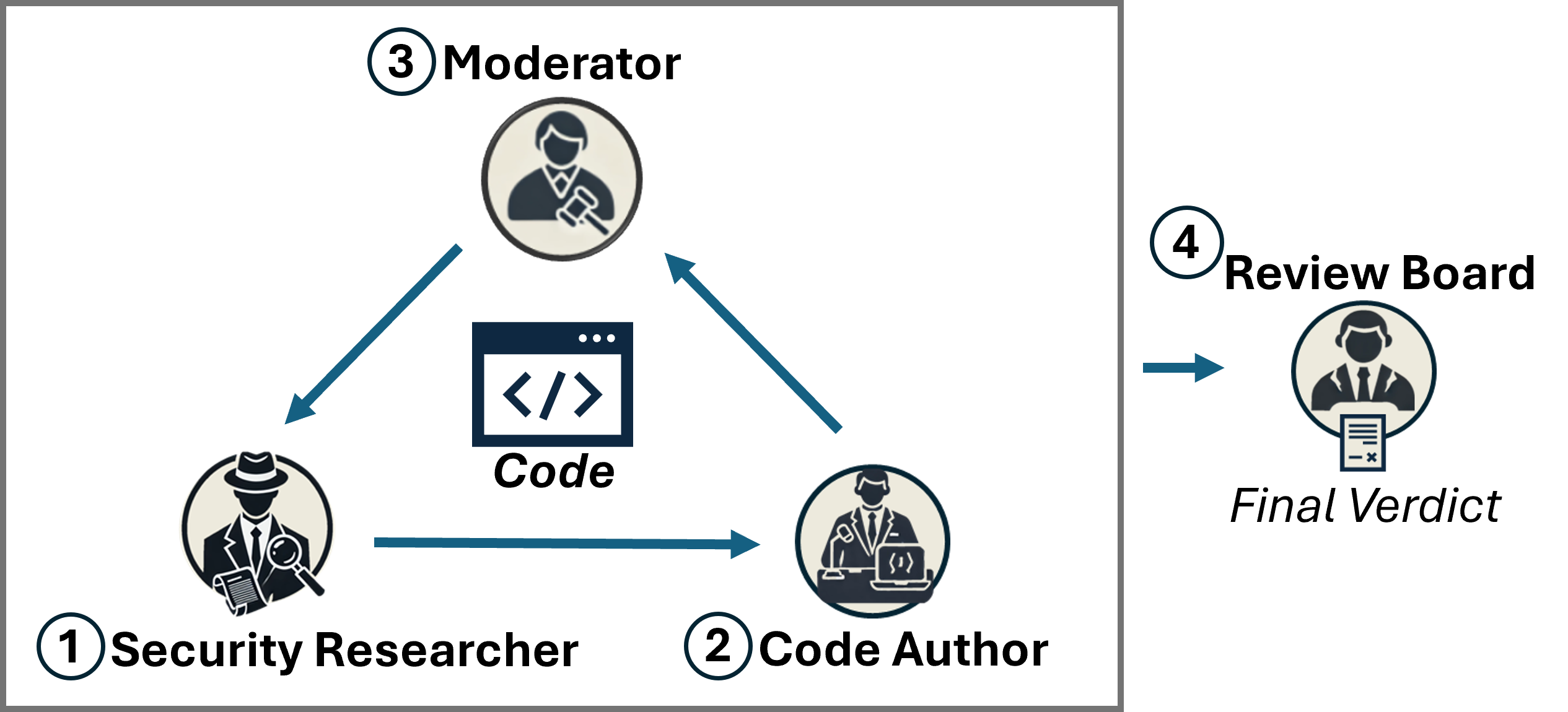}
    \caption{\tool{} architecture which uses a courtroom-inspired multi-agent framework where the \emph{security researcher} (prosecutor) identifies potential vulnerabilities, the \emph{code author} (defense attorney) provides counterarguments, the \emph{moderator} (judge) summarizes key points, and the \emph{review board} (jury) delivers the final verdict on code vulnerabilities. 
    }
    \label{fig:architecture}
    \vspace{-0.6cm}
\end{figure}

\tool{} architecture is shown in Fig.~\ref{fig:architecture}, with an example output in Fig.~\ref{fig:example}.
The first agent acts as a \emph{security researcher} and is responsible to analyze the given code and identify potential vulnerabilities. 
The second agent, namely the \emph{code author agent}, aims to provide counterarguments and justifications to refute the claims made by the \emph{security researcher agent}.
The third agent, which is the \emph{moderator agent}, is tasked with summarizing the arguments presented by both the \emph{security researcher} and the \emph{code author agents} to ensure a structured exchange of viewpoints between them. 
\tool{} supports multiple rounds of interaction among the first three agents before passing their viewpoints to the last agent. The last agent, namely the \emph{review board agent}, is tasked with analyzing all the information given by the \emph{moderator}, the \emph{security researcher}, and the \emph{code author agents}.
Based on this collective information, the \emph{review board} delivers a final verdict regarding the security status of the code. 
We chose to have the agents simulate a jury over a judge sentencing scenario, as previous studies have shown that jury participation can help prevent excessively harsh punishments~\cite{hans2015death,mcleod2024democratic}. 
The details of the agent's interaction will be discussed in the following subsections. The prompts can be found in our online appendix~\cite{ourreplicationpackageall}.

\subsection{Security Researcher Agent (Prosecutor Role)}
In \tool{}, the \emph{security researcher agent} functions similarly to a prosecutor in a courtroom. A prosecutor’s role is to investigate evidence, establish the facts, and build a case against a defendant (i.e., the party that is accused in a court of law). In the case of vulnerability detection, the \emph{security researcher agent} inspects the provided code, identifies potential vulnerabilities, and presents a rationale for why these vulnerabilities may pose a threat. An example is shown in Fig.~\ref{fig:example}, where it highlights two potential vulnerabilities in the code, along with their reasoning and impact.

In the first round, the agent receives a code function and determines whether the code is vulnerable. In subsequent rounds, it also receives the output generated by two other agents (i.e., \emph{code author} and \emph{moderator}). 
Beyond simply reporting any vulnerability, the agent is explicitly asked to include (1) a description of each vulnerability, (2) the reasoning behind it, and (3) the potential impact if exploited. 
We require the agent to include its reasoning, as previous studies~\cite{wei2022chain,kojima2022large,cheng2024chainlm} show that prompting LLMs to articulate their reasoning can significantly enhance their performance. 
Furthermore, including the potential impact of each vulnerability gives the other agents a clearer perspective when they form their answer.

\subsection{Code Author Agent (Defense Attorney Role)}

The \emph{code author agent} in \tool{} acts similarly to a defense attorney in a courtroom scenario. 
While the prosecutor aims to prove that the defendant is guilty, the defense attorney counters these allegations by advocating for the defendant's innocence. 
Applying this scenario to vulnerability detection, the \emph{code author agent} seeks to defend the integrity and security of their code, providing arguments and evidence to counter the \emph{security researcher agent's} claims. If the vulnerability is valid, the agent can acknowledge the issue and propose effective mitigation strategies.

The prompt provides the agent with the code snippet under investigation, along with information generated by the \emph{security researcher agent}, which outlines a list of potential vulnerabilities~\cite{ourreplicationpackageall}. The agent is then tasked with clearly articulating the following outputs: (1) identification of the vulnerability being discussed, (2) explicit acknowledgment or refutation of the vulnerability, and (3) detailed reasoning justifying the refutation or mitigation methods in case the vulnerability is acknowledged. 
An example of the \emph{code author agent's} output is also shown in Fig.~\ref{fig:example}, where it refutes one vulnerability and provides reasoning for it.

Since the \emph{security researcher agent} may highlight multiple potential vulnerabilities, it is crucial for the \emph{code author agent} to clearly specify which vulnerability it addresses in each response. 
Explanations from the \emph{code author agent} enable other agents (the \emph{moderator} and \emph{review board}) to better understand the justification or rationale behind the agent's position. For the \emph{security researcher} agent, this reasoning would be used for building their rebuttal. 
This structured approach helps ensure that subsequent agents are equipped to offer well-informed counterarguments, summaries, or final evaluations.

\subsection{Moderator Agent (Judge Role)}
Just as judges provide guidance to jurors and uphold fairness in a courtroom scenario, within the context of vulnerability detection, this agent is tasked with objectively summarizing the arguments presented by both the \emph{security researcher agent} (prosecutor) and the \emph{code author agent} (defense attorney). 
It should focus only on the facts, avoiding any bias toward either side. 
This impartial summary helps the \emph{review board agent} in reaching a decision, while also clarifying critical discussion points for subsequent rounds of argumentation by the \emph{security researcher} and \emph{code author agents}.

\begin{figure}[t]
    \centering
    \includegraphics[width=\linewidth]{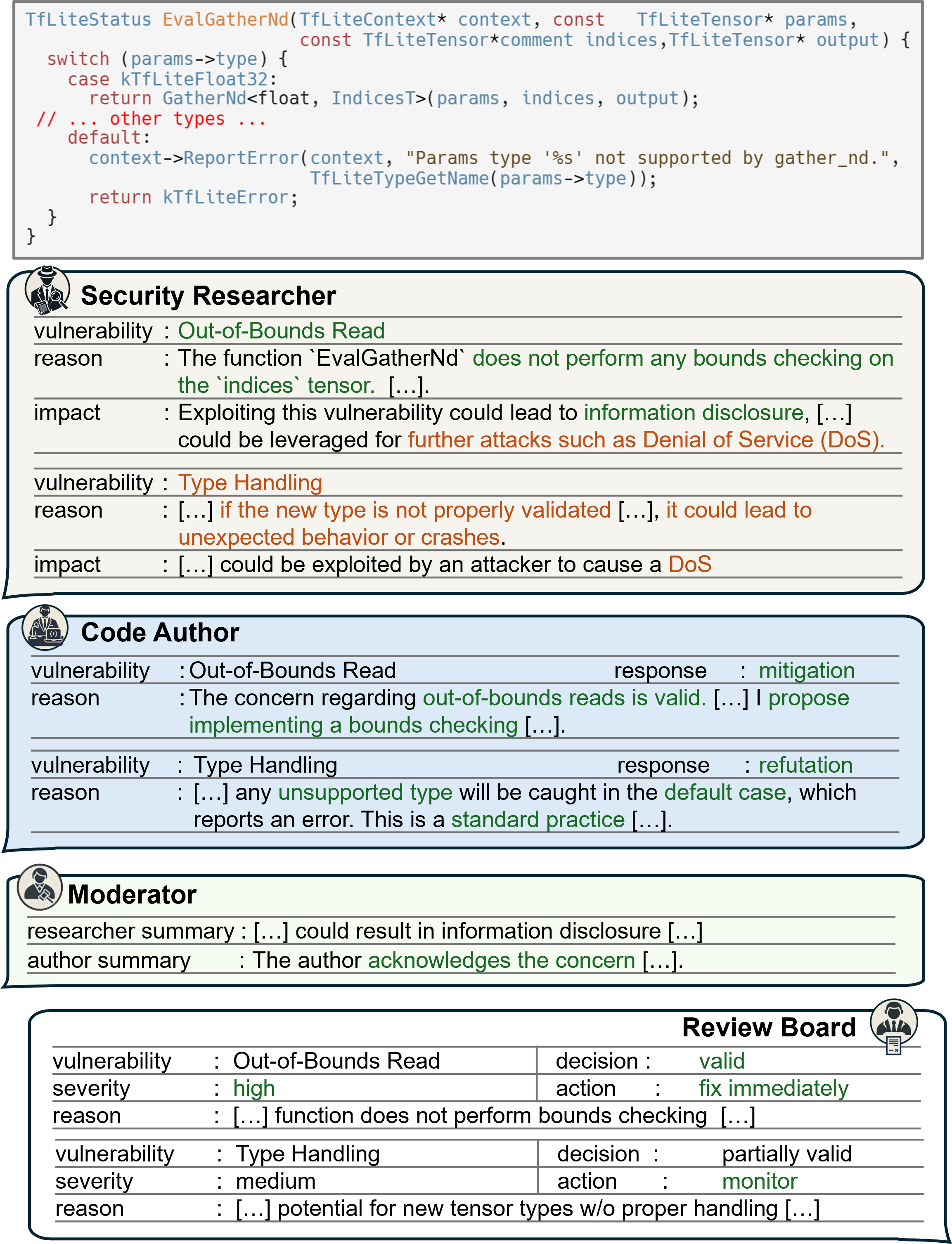}
    \caption{Example of vulnerable code (CVE-2021-37687) and analysis results from \tool{}. \textcolor{OliveGreen}{Green} text indicates correct arguments, while \textcolor{Bittersweet}{orange} text indicates misleading or exaggerated claims. The code is vulnerable as it does not check for negative indices, allowing an attacker to read arbitrary data by crafting a model with negative values in indices.}
    \label{fig:example}
    \vspace{-0.4cm}
\end{figure}

In addition to the code snippet, the \emph{moderator} receives the \emph{security researcher}’s identified vulnerabilities and the \emph{code author}’s counterarguments.  
The \emph{moderator}’s output consists of concise summaries of these two viewpoints, highlighting the key points from each. 
Fig.~\ref{fig:example} shows an example of the \emph{moderator agent} summarizing the arguments from the \emph{security researcher} (e.g., the number of potential vulnerabilities identified, their main reasons, and impacts) as well as from the \emph{code author}, such as which vulnerabilities were acknowledged or refuted with the main reasoning.

\subsection{Review Board (Jury Role)}
In a traditional courtroom, the jury weighs all evidence and testimony presented during a trial, alongside any guidance from the judge, to determine whether a defendant is guilty or not.  Similarly, in vulnerability detection, the \emph{review board agent} is tasked to evaluate the code’s security status based on the evidence presented by both the \emph{security researcher} and \emph{code author agents}, as well as the \emph{moderator}’s summary. Its purpose is to deliver a final verdict on whether the potential vulnerabilities identified by the \emph{security researcher} are indeed valid.  
This agent receives (1) the summary from the \emph{moderator} agent; (2) the list of potential vulnerabilities from the \emph{security researcher} agent; (3) the \emph{code author} agent’s counterarguments; (4) the original code snippet. 

The \emph{review board agent} is tasked with generating several outputs.
First, it must extract the names of potential vulnerabilities from the list provided by the \emph{security researcher} and proceed with a decision for each one. 
Second, the agent must determine whether each identified vulnerability is ``valid,'' ``invalid,'' or  ``partially valid.'' This process is similar to a jury deciding whether a defendant is guilty or not guilty. Third, the agent must assess the severity of the vulnerability, similar to how a jury determines the extent of damages in a trial. However, instead of assigning an open-ended value, the \emph{review board} needs to categorize severity as ``low,'' ``medium,'' or ``high.'' If the vulnerability is deemed invalid, the severity is marked as ``none.'' 
Fourth, the agent must provide a recommended action based on the decision and severity level. For example, if a vulnerability is deemed valid and has a high severity, the recommended action should be ``fix immediately.’’ Finally, the agent must generate an explanation for its decisions. Unlike a real jury, we require the \emph{review board agent} to provide reasoning, as it improves LLMs' performance~\cite{wei2022chain,kojima2022large,cheng2024chainlm} and helps users understand the decision.

We use a threshold rather than a simple binary ``vulnerable or not,'' because it offers flexibility for users to tailor vulnerability detection to their tolerance for risk and false positives. For instance, if a user prefers to catch even the smallest chance of vulnerability, they can opt to flag any partially valid vulnerabilities. 
Conversely, if a user wants fewer false positives, they can raise the threshold to consider only higher-severity issues.

We flagged a function as vulnerable if at least one vulnerability raised by the \emph{security researcher agent} is deemed valid at high severity with recommended action fix immediately by the \emph{review board agent}. This threshold ensures that only the critical and well-substantiated vulnerabilities contribute to the classification, reducing false positives from lower-risk or ambiguous issues. We aim to reduce noise by excluding vulnerabilities marked as partially valid or those with lower severity levels. Fig.~\ref{fig:example} also includes an example of the review board agent's output. It classifies the first vulnerability (Out-of-Bounds Read) as ``valid'' with ``high'' severity. Meanwhile, the second vulnerability (Type Handling) is marked as ``partially valid'' with ``medium'' severity. As at least one confirmed vulnerability is present, the final verdict for this code is vulnerable.

\subsection{Agent Interaction}
In \tool{}, four agents interact sequentially: the \emph{security researcher agent}, the \emph{code author agent}, the \emph{moderator agent}, and finally the \emph{review board agent}. During the first round, the \emph{security researcher agent} (1) has access only to the original code snippet. In subsequent rounds, it also receives information from the \emph{code author} and \emph{moderator}. Similarly, the \emph{code author agent} (2) initially sees only the \emph{security researcher agent’s} findings, but in the next rounds, it can also incorporate the \emph{moderator}’s (3) output. This setup is similar to a courtroom process, which can allow for multiple rounds of arguments and counterarguments between the prosecutor, defense attorney, and judge.
The \emph{review board agent} (4), 
mirroring a jury role, provides its verdict only after all discussions have concluded. It does not participate in the debate itself but bases its final decision on the evidence and summaries presented by the other three agents.

Allowing multiple discussion rounds may enhance the iterative refinement of analyses and arguments. It also provides both the \emph{security researcher} and \emph{code author agents} with more opportunities for rebuttal. 
In this study, we initially limit the number of discussion rounds to one (Section~\ref{sec:rq1_result}) and then assess how changing the number of rounds affects \tool{}’s performance (Section~\ref{sec:rq3_result}). 

\section{Experimental Setup}\label{sec:setting}
\subsection{Dataset}
We utilized the PrimeVul dataset, recently proposed by Ding et al.~\cite{ding2024vulnerability}, to evaluate vulnerability detection approaches. Specifically, we used the PrimeVul pair dataset, which presents a challenging benchmark, as the vulnerable and benign code functions share at least 80\% of their string content. 
The pair dataset includes 3,789 training pairs, 480 validation pairs, and 435 test pairs, with no data leakage across splits.
In contrast, the full PrimeVul dataset contains 175,797 training functions (4,862 vulnerable and 170,935 benign) and 23,948 validation functions (593 vulnerable and 23,355 benign).

\subsection{Research Questions}
This study aims to answer the following research questions (RQs):

{\bf RQ1: How does the performance of \tool{} compare to that of single-agent and multi-agent baselines on PrimeVul?} 
Prior work~\cite{ding2024vulnerability} demonstrated the poor performance of GPT-3.5 and GPT-4 for vulnerability detection on pairwise vulnerable and benign functions in PrimeVul. These findings highlight the limitations of single-agent prompting strategies for this challenging setting. 
In this RQ, we investigate whether \tool{}, a multi-agent LLM-based framework (Section~\ref{sec:approach}), can outperform both single-agent methods~\cite{ding2024vulnerability} and existing multi-agent approaches such as GPTLens~\cite{hu2023large} on the PrimeVul pair.

{\bf RQ2: How does instruction tuning on role-specific agents influence the performance of automated vulnerability detection?}
\tool{} consists of four agents, each with a specific role. Instruction tuning can be applied to each agent to strengthen its performance in its designated task. In this RQ, we investigate whether this role-specific instruction tuning can enhance the overall performance of \tool{}.

{\bf RQ3: Does allowing multiple rounds of debate  among role-specific agents in \tool{} lead to improvement compared to a single-pass analysis?}
In previous RQs, we focused on conducting a single round of discussions among agents to reach a final verdict. However, allowing multiple rounds of discussion could give agents the opportunity to refine their arguments to reach a more accurate decision. In this RQ, we examine whether enabling multiple rounds of discussion can enhance \tool{}'s performance.

{\bf RQ4: How does each agent contribute to the overall performance of \tool{} }
\tool{} features four distinct roles, \emph{security researcher}, \emph{code author}, \emph{moderator}, and \emph{review board}, which is designed to follow a courtroom-style scenario. While the results from previous RQs highlight \tool{}’s effectiveness, it is not known whether each agent is essential to the final prediction and how it impacts the performance. 
In this RQ, we do the ablation study that omits one agent at a time to verify each agent’s contribution.

\subsection{Baseline and Models}\label{sec:model}
\noindent{\bf \textit{Baselines.}} 
We employed both DL-based and LLM-based vulnerability detection techniques as single-agent baselines. For the DL-based methods, we used several state-of-the-art models widely adopted in prior vulnerability detection studies~\cite{ding2024vulnerability, xiong2024vuld, zhou2024large, chen2023diversevul}:
\begin{itemize}[leftmargin=0.05\linewidth, itemsep=1pt, parsep=0pt, topsep=1pt, partopsep=0pt]
    \item CodeBERT~\cite{feng2020codebert}: A bimodal pre-trained transformer model trained on source code and natural language.
    \item CodeT5~\cite{wang2021codet5}: A transformer model that learns to understand and generate code across multiple programming languages.
    \item UniXCoder~\cite{guo2022unixcoder}: A cross-modal pre-trained model that jointly learns representations of code, ASTs, and execution traces. 
    \item LineVul~\cite{fu2022linevul}: A line-level transformer-based vulnerability prediction model detecting vulnerable lines through fine-grained context representations. 
\end{itemize}
For CodeBERT, CodeT5, and UniXCoder, we performed two separate fine-tuning: (1) using the full PrimeVul training and validation sets, and (2) using only the pair-wise subsets from PrimeVul. For LineVul, we first evaluated the model provided in the previous study~\cite{fu2022linevul}. In addition, we further fine-tuned LineVul using the pair-wise subset to ensure a consistent comparison with the other DL models. For the LLM-based approach, we utilized the chain of thought (CoT)~\cite{wei2022chain} prompting technique, which achieved the highest results in the PrimeVul~\cite{ding2024vulnerability}. 
CoT prompting encourages LLMs to reason through problems step by step to arrive at an answer. 

For the multi-agent baseline, we utilized an approach from a recent study, GPTLens~\cite{hu2023large}. GPTLens operates with two types of agents: \emph{auditor agent}, which identifies potential vulnerabilities, and a \emph{critic agent}, which evaluates and scores the identified vulnerabilities. The final decision is based on the critic's score; if the score is less than 5, it is classified as benign; otherwise, it is considered vulnerable. There are several key differences between GPTLens and \tool{}. \tool{} follows a courtroom-inspired setting with four role-specific agents, whereas GPTLens relies on a two-agent structure. Unlike GPTLens, where the \emph{critic agent} determines the final decision, \tool{} used a different agent (i.e., \emph{review board agent}) for this task, designed to be neutral toward the \emph{security researcher} and \emph{code author agents}. Recent studies~\cite{gao2024large, hu2024unlocking} suggest that employing domain-specific multi-agent systems that reflect a more realistic scenario leads to better decision making, further supporting our approach in \tool{}.

\vspace{0.2cm}\noindent{\bf \textit{Models.}} 
We used well-established OpenAI models known for their strong performance, as confirmed by prior research~\cite{widyasari2024beyond,hu2023large,zhou2024large_emerging}. 

{\bf GPT-3.5~\cite{openai2023gpt35}}: GPT-3.5 is one of the first widely popular OpenAI LLM models. Following PrimeVul~\cite{ding2024vulnerability}, we also evaluate the performance of this model in our study. In our evaluation, we utilized \emph{gpt-3.5-turbo-0125}, the latest iteration of the GPT 3.5 model with a context window of 16k tokens. 

{\bf GPT-4o~\cite{openai2024gpt4o}}: 
We chose GPT-4o, one of OpenAI’s flagship models, over GPT-4~\cite{openai2023gpt4} (used in PrimeVul’s evaluation) because OpenAI indicates that GPT-4o offers performance comparable to GPT-4 while significantly reducing computational costs~\cite{openai2024gpt4o}. 
Specifically, we used \emph{gpt-4o-2024-08-06} allows up to 128k tokens in its context window and 16k output tokens.

\subsection{Evaluation Metrics}\label{sec:metrics}
Following the approach in PrimeVul~\cite{ding2024vulnerability}, we adopt a pair-wise evaluation strategy. It is designed to assess the model's prediction on pairs of vulnerable and benign code, rather than the single codes. 
This method highlights the model’s capacity to differentiate subtle variations that make one function vulnerable and the other benign. We report five metrics. Pair-wise Correct Prediction (P-C) records pairs where both functions are classified correctly. Pair-wise Vulnerable Prediction (P-V) records pairs where both functions are incorrectly labeled as vulnerable, while Pair-wise Benign Prediction (P-B) records pairs labeled incorrectly as benign. Pair-wise Reversed Prediction (P-R) captures cases where the model assigns opposite labels to the pair, predicting ``benign'' for the ``vulnerable'' code and ``vulnerable'' for the ``benign'' one. Pair-wise Error (E) reflects pairs in which one function is misclassified. Among these metrics, P-C is our primary evaluation metric. A high P-C indicates accurate vulnerability detection that avoids false positives and negatives, making it a more robust indicator of overall performance.

To provide more comprehensive discussion in \tool{} and baselines comparison, we also report standard classification metrics: Precision (P), Recall (R), and False Positive Rate (FPR), following the formula used in the previous study~\cite{ding2024vulnerability}. 
These traditional metrics provide additional insights into the model’s general classification behavior, particularly in individual predictions.

\subsection{Methods}
\subsubsection{Answering RQ1:} 
We compare \tool{} with both single-agent (CodeBERT~\cite{feng2020codebert}, CodeT5~\cite{wang2021codet5}, UniXCoder~\cite{guo2022unixcoder}, LineVul~\cite{fu2022linevul}, and Ding et al.'s CoT\cite{ding2024vulnerability}) and multi-agent (GPTLens) baselines. For the LLM-based techniques, we utilize two different LLMs, i.e., GPT-3.5 and GPT-4o (see Section~\ref{sec:model}). 
We evaluate each approach using several metrics (P-C, P-V, P-B, P-R, Error, P, R, and FPR). 

\subsubsection{Answering RQ2:}\label{sec:rq2method}  For this RQ, we conducted instruction tuning using samples drawn from GPT-4o’s validation set. After running GPT-4o on the set, we collected its correct outputs to build an instruction-tuning dataset for a GPT-4o model. We focus on GPT-4o because, as shown in RQ1, \tool{} with GPT-4o outperforms both single-agent and multi-agent baselines and achieves the best overall results.
Our goal is to determine whether instruction tuning can further enhance its performance.
We chose GPT-4o for generating the instruction data to ensure consistency in style and reasoning patterns. 
By using the same models for generating data, we reduce the risk of mismatches that can occur when instruction tuning with outputs from a completely different model.

For instruction tuning data, we use 50 code pairs (vulnerable and benign) with the fewest tokens in the validation set to help manage the tuning cost. We chose 50 samples based on preliminary experiments comparing different dataset sizes (25, 50, and 75 pairs), as 50 resulted in significant performance gains over 25 samples, while achieving comparable results to using 75 samples~\cite{ourreplicationpackageall}. In addition, OpenAI has highlighted that GPT-4o can achieve strong results with ``as few as a few dozen examples''~\cite{openai2024gpt4o_finetuning}. When creating our instruction tuning dataset, we initially included the true label in the prompt, but then removed direct label disclosures in the final dataset. For example, for the code author agent~\cite{ourreplicationpackageall}, we included vulnerability-related information such as CVE ID and label. Similarly, for the security researcher agent, we added the CVE description and label for both vulnerable and benign code, prompting the agent to connect code to relevant vulnerabilities. For the review board agent, the prompt included ground-truth labels but also emphasized that the agent should not reference them explicitly in its reasoning. In contrast, the moderator agent’s prompt excluded any ground-truth vulnerability information and was cleaned to avoid indirect label leakage from chat history. As a result, each prompt retained its original role-specific structure and was paired with the correct assistant output.

For this instruction tuning data, we ensure that the agent’s predicted label (``vulnerable'' or ``benign'') matches the ground-truth label. Additionally, we manually verify the quality of the agent’s explanation, confirming that it correctly identifies the presence or absence of a vulnerability. An example of the vulnerable instance is shown in the online appendix~\cite{ourreplicationpackageall}; this vulnerability is classified as CWE-416: Use After Free. The security researcher agent correctly identifies the issue within the CleanWriters function, where the sequence of operations may lead to undefined behavior if \texttt{gf\_isom\_box\_del()} uses the writer object after it has been freed. The explanation aligns well with the root cause described in CVE-2020-35980.
For the corresponding benign variant~\cite{ourreplicationpackageall}, the \emph{code author agent} correctly refutes the presence of a vulnerability by noting that \texttt{gf\_list\_del\_item()} only removes the items from the list without deallocating their memory. The justification provided by the \emph{review board agent} also matches the \emph{code author's agent} clarification and supports the benign ground-truth label.

We evaluated eight scenarios: (1) no instruction tuning; (2-5) tuning only the \emph{security researcher}, \emph{code author}, \emph{moderator}, or \emph{review board agent}; (6) tuning \emph{security researcher} and \emph{code author}; (7) tuning \emph{security researcher}, \emph{code author}, and \emph{moderator}; and (8) tuning all agents.
To systematically assess the impact of instruction tuning, we first applied it to individual agents to understand their specific contributions. Then, we incrementally introduced instruction tuning in the order of agent interaction, allowing us to analyze how tuning at different stages influences the overall performance of \tool{}. This stepwise approach provides insights into the cumulative effects of instruction tuning across multiple agents.

\subsubsection{Answering RQ3:} We experimented with three values of $k$ (i.e., the number of interaction rounds): 1, 2, and 3 in \tool{}. We selected these values to systematically investigate the impact of multiple interaction rounds while maintaining computational feasibility. For the base model, we used GPT-4o with an instruction-tuned agent, as it yielded the strongest overall performance. 

\subsubsection{Answering RQ4:} 
In this RQ, we conduct an ablation study by systematically removing each agent to assess its necessity. Since the \emph{security researcher agent} is fundamental for identifying potential vulnerabilities, we retain it in all scenarios. Additionally, since the moderator primarily provides summaries rather than making explicit decisions on vulnerabilities, we do not include a scenario where the moderator is responsible for delivering the final verdict.

We evaluated four scenarios: (1) the full \tool{} (baseline) includes all four agents; (2) removing the \emph{moderator}; (3) removing the \emph{code author}; (4) removing two agents simultaneously, using either (a) only the \emph{security researcher} and \emph{review board} or (b) only the \emph{security researcher} and \emph{code author}.
For scenario 4(b), as the final decision is made by the \emph{code author agent}, we classify the code as vulnerable if it mitigates at least one of the potential vulnerabilities.

By comparing performance across these scenarios, we can identify which agents play the most critical roles. For example, does a \emph{moderator} assist the \emph{review board} in reaching a final decision, and how essential is a different agent for delivering verdicts? This ablation would highlight the necessity of each agent in \tool{}.

\subsection{Implementation Details}
We built \tool{} on top of AgentVerse~\cite{chen2024agentverse}, a recently developed framework that simplifies the deployment of multiple LLM-based agents. For single-agent baselines, we used the prompt in PrimeVul~\cite{ding2024vulnerability}, while for GPTLens~\cite{hu2023large}, we employed the provided replication package~\cite{gptlens}. Because the original GPTLens study focused on smart contracts, we adapted its prompt by replacing references to ``smart contract'' with ``code'' to generalize the use case. We used the best settings reported in their study, including two auditor agents and a top-$k$ value of 3. We counted the results as vulnerable if they got a final score of at least 5 (out of 9). 
All GPT-based models were accessed via the OpenAI API, which was also used to generate instruction tuning data. For the fine-tuning itself, we used OpenAI’s built-in dashboard feature for custom GPT-3.5 and GPT-4o instruction tuning~\cite{openai2024gpt4o_finetuning}.
To make the results more deterministic, we set the temperature to 0 for both our single-agent baselines and \tool{}. For GPTLens~\cite{hu2023large}, we followed the same setting used in its evaluation (0.7 for the auditor agent and 0 for the critic agent).

\section{Results}\label{sec:results}
\subsection{RQ1: \tool{} vs. Baselines}\label{sec:rq1_result}
Table~\ref{tab:rq1_res} presents the performance of \tool{} against both single-agent (CodeBERT~\cite{feng2020codebert}, CodeT5~\cite{wang2021codet5}, UniXCoder~\cite{guo2022unixcoder}, LineVul~\cite{fu2022linevul}, and Ding et al.'s CoT prompts~\cite{ding2024vulnerability}) and multi-agent (GPTLens~\cite{hu2023large}) baselines. 
In addition to the pair-wise metrics~\cite{ding2024vulnerability} (P-C, P-V, P-B, P-R, Error), we also report traditional metrics (P, R, FPR).

\begin{table*}[]
\begin{threeparttable}
\caption{Results of various vulnerability detection methods on PrimeVul paired functions. An upward-pointing arrow ($\uparrow$) indicates that a higher score is better, while a downward-pointing arrow ($\downarrow$) indicates that a lower score is better. 
}\label{tab:rq1_res}
\setlength{\tabcolsep}{0.75em}
\begin{tabular}{|l|l|l|r|r|r|r|r|r|r|r|}
\hline
\rowcolor[HTML]{C0C0C0} 
\multicolumn{1}{|c|}{\cellcolor[HTML]{C0C0C0}\textbf{Model}} & \textbf{Agents}       & \multicolumn{1}{c|}{\cellcolor[HTML]{C0C0C0}\textbf{Method}} & \multicolumn{1}{c|}{\cellcolor[HTML]{C0C0C0}\textbf{P-C $\uparrow$}} & \multicolumn{1}{c|}{\cellcolor[HTML]{C0C0C0}\textbf{P-V $\downarrow$}} & \multicolumn{1}{c|}{\cellcolor[HTML]{C0C0C0}\textbf{P-B $\downarrow$}} & \multicolumn{1}{c|}{\cellcolor[HTML]{C0C0C0}\textbf{P-R $\downarrow$}} & \multicolumn{1}{c|}{\cellcolor[HTML]{C0C0C0}\textbf{Error $\downarrow$}}& \multicolumn{1}{c|}{\cellcolor[HTML]{C0C0C0}\textbf{P $\uparrow$}}& \multicolumn{1}{c|}{\cellcolor[HTML]{C0C0C0}\textbf{R $\uparrow$}}& \multicolumn{1}{c|}{\cellcolor[HTML]{C0C0C0}\textbf{FPR $\downarrow$}}\\ \hline
\multirow{-1}{*}{CodeBERT~\cite{feng2020codebert}}  & \multirow{2}{*}{Single-Agent} & Fine-tuned full$^\dagger$~\cite{ding2024vulnerability} & 5 & 37 & 386 & 7 & 430 & 0.49 & 0.10 & 0.10  \\  \cline{3-11} 
    &                              & Fine-tuned pair$^\ddagger$ &  0 & 435 & 0 & 0 & 435 & 0.50 & 1.00 & 1.00   \\ \cline{1-1} \cline{2-11} 
\multirow{-1}{*}{CodeT5~\cite{wang2021codet5}}  & \multirow{2}{*}{Single-Agent} & Fine-tuned full$^\dagger$~\cite{ding2024vulnerability} & 0 & 57 & 373 & 5 & 435 & 0.48 & 0.13 & 0.14  \\  \cline{3-11} 
    &                              & Fine-tuned pair$^\ddagger$ & 0 & 434 & 1 & 0 & 435 & 0.50 & 0.99 & 0.99     \\ \cline{1-1} \cline{2-11}  
\multirow{-1}{*}{UniXCoder~\cite{guo2022unixcoder}}  & \multirow{2}{*}{Single-Agent} & Fine-tuned full$^\dagger$~\cite{ding2024vulnerability} & 4 & 26 & 403 & 2 & 431 & 0.52 & 0.07 & 0.06  \\  \cline{3-11} 
    &                              & Fine-tuned pair$^\ddagger$ & 6 & 412 & 17 & 0 & 429 & 0.50 & 0.96 & 0.95   \\ \cline{1-1} \cline{2-11}  
\multirow{-1}{*}{LineVul~\cite{fu2022linevul}}  & \multirow{2}{*}{Single-Agent} & Fine-tuned$^\S$~\cite{fu2022linevul}  & 7 & 89 & 336 & 3 & 428 & 0.51 & 0.22 & 0.21 \\  \cline{3-11} 
    &                              & Fine-tuned pair$^\ddagger$ &  0 & 434 & 1 & 0 & 435 & 0.50 & 0.10 & 1.00   \\ \hline 
       &   Single-Agent  & Ding et al.'s CoT~\cite{ding2024vulnerability} & 18 & 16 & 381 & 20 & 417 & 0.49 & 0.08 & 0.08          \\ \cline{2-11} 
       &     & GPTLens~\cite{hu2023large}         & 20 & 388 & 3 & 24 & 415 & 0.50 & 0.94 & 0.95        \\ \cline{3-11} 
\multirow{-3}{*}{GPT-3.5}         & \multirow{-2}{*}{Multi-Agent}  & \tool{}  &   \textbf{68} & 40 & 265 & 62 & 367  & 0.51 & 0.25 &  0.23        
\\ \hline
       & Single-Agent          & Ding et al.'s CoT~\cite{ding2024vulnerability}    & 40 & 43 & 323 & 29 & 395 & 0.53 & 0.19 & 0.17        \\ \cline{2-11} 
       &     & GPTLens~\cite{hu2023large}         & 44 & 241 & 122 & 28  & 391 &  0.51 & 0.65 & 0.62        \\ \cline{3-11}  
\multirow{-3}{*}{GPT-4o}          & \multirow{-2}{*}{Multi-Agent}  & \tool{}        &  \textbf{81} & 179 & 125 & 50 & 354 & 0.53 & 0.60 & 0.52    
\\ \hline

\end{tabular}
\begin{tablenotes}
  \small
  \item \begin{minipage}[t]{0.48\linewidth} $\dagger$ Fine-tuned on full PrimeVul train and val sets (4,862 vuln / 170,935 benign). \end{minipage}
  \hfill
  \begin{minipage}[t]{0.48\linewidth} $\ddagger$ Fine-tuned on paired PrimeVul train and val sets (3,789 vuln / 3,789 benign).  \end{minipage}
  \item $\S$ Fine-tuned model from prior work~\cite{fu2022linevul}.
\end{tablenotes}
\end{threeparttable}
\vspace{-0.8em}
\end{table*}

\paragraph{\textbf{Deep learning-based Method.}} Our results indicate that DL models trained on the balanced pair dataset fail to effectively differentiate between vulnerable and benign functions. While these models achieve nearly perfect recall (R=0.96-1.00), it comes at the cost of maximum false positive rates (FPR=0.99-1.00). This leads to zero or near-zero correct predictions for both functions in a pair (P-C=0-6). When trained on the full dataset (imbalanced), DL models slightly improve, reducing FPR to approximately 0.10-0.14; however, they still struggle significantly, with very few correct pair predictions (P-C$\le$7) and low recall (R$\le$0.13). Most functions are incorrectly labeled benign (P-B=336-403), indicating that the models cannot pick up subtle vulnerability patterns amid class imbalance, so they miss most of the real vulnerability.

These results highlight the limitations of traditional DL models in capturing subtle vulnerability patterns at the function-pair level. While DL-based models are more lightweight and computationally efficient, our results demonstrate that this efficiency comes at a cost in actionable performance. For example, compared to the highest performing DL baseline, LineVul, achieves only 7 for P-C, with a modest FPR of 0.21 and a recall of 0.22. \tool{} (using GPT-3.5) achieves a P-C of 68 (nearly 10× higher) while maintaining a comparable FPR and recall of 0.23 and 0.25. With GPT-4o, \tool{} pushes the P-C even further to 81 (12× higher than LineVul), with recall reaching 0.60. Although this comes at the cost of a higher FPR (0.52), the trade-off results in vastly more correct pairwise classifications and a significantly higher yield of true vulnerabilities.

\paragraph{\textbf{LLM-based Method (GPT-3.5).}} The single-agent CoT prompt obtains modest accuracy (P-C=18, FPR=0.08) but misses most vulnerabilities (R=0.08). Meanwhile, GPTLens boosts recall to 0.94; however, it comes with a high FPR of 0.95 with a small number of correct pairs (20). \tool{} obtains 68 correct pairs, which is 3.8x the CoT baseline and 3.4x GPTLens, while keeping FPR to 0.23. 

GPTLens is configured to output the top-3 potential vulnerabilities per function, following the best setup from the original study~\cite{hu2023large}. Since the \emph{auditor agent} runs twice, this results in up to six potential vulnerabilities per function. Indeed, most \emph{auditor} outputs list six vulnerabilities, with an average of 5.6 and a median of 6. 
Meanwhile, \tool{} produces an average of 1.74 vulnerabilities per function, with a median of 2. In other words, \tool{} produces fewer vulnerabilities compared to GPTLens, but it has more precise vulnerability detection, shown by a higher P-C.

\paragraph{\textbf{LLM-based Method: GPT-4o.}} 
\tool{} achieves the highest number of pair-wise correct predictions (P-C=81), outperforming both the CoT baseline (P-C=40) and GPTLens (P-C=44). This is 2.03× the performance of the single-agent baseline and 1.84× that of GPTLens. These gains highlight \tool{}'s effectiveness at correctly identifying both the vulnerable and benign functions within a pair that have subtle differences.

In terms of traditional classification metrics, \tool{} achieves a precision of 0.53, on par with the CoT baseline (0.53) and slightly above GPTLens (0.51). Its recall is 0.60, lower than GPTLens (0.65), but significantly higher than CoT (0.19). The FPR for \tool{} is 0.52, which is lower than GPTLens (0.62) but higher than CoT (0.17). This trade-off illustrates \tool{}’s balanced behavior; it identifies more true vulnerabilities than CoT while maintaining a lower FPR than GPTLens. In terms of P-B, which measures how often both functions in the pair were incorrectly classified as benign, \tool{} (P-B=125) is significantly lower than the CoT baseline (P-B=323) and comparable to GPTLens (P-B=122).  
Despite these improvements, \tool{} does not outperform the alternatives on every metric. GPTLens still achieves the highest recall, making it preferable in scenarios where catching as many vulnerabilities as possible is critical.  

In the GPTLens experiment using GPT-4o, the first agent highlights a similar number of potential vulnerabilities as it did under GPT-3.5, with a median of 6 and a mean of 5.76. Meanwhile, \tool{} with GPT-4o reports fewer vulnerabilities (median 4, mean 4.11) compared to GPTLens. Despite highlighting fewer vulnerabilities, \tool{} achieved higher P-C and lower P-B scores than GPTLens.  
This suggests that the additional agents in \tool{}, can effectively distinguish between justified and unjustified vulnerability claims, leading to more accurate final predictions.

Our qualitative analysis further illustrates how \tool{}’s structured, role-specific dialogue improves both fidelity and precision of vulnerability detection. Each agent in \tool{} has a specific role. The \emph{security researcher agent} analyzes the function to identify vulnerabilities. For example, in Fig.~\ref{fig:example}, the \emph{security researcher} highlights two vulnerabilities: (1) an Out-of-Bounds Read due to unvalidated negative indices (CWE-125), and (2) a Type Handling issue. The first is a valid case, while the second reflects an overcautious concern since the existing code already handles unsupported types safely.  
After reviewing these findings, the \emph{code author agent} agrees with the Out-of-Bounds Read assessment but refutes the Type Handling concern.  
The \emph{moderator agent} then summarizes key points from both agents, which aim to help the \emph{review board agent} in making the decision.  
Based on this information, the \emph{review board agent} concludes that the Out-of-Bounds Read is a genuine vulnerability.  
Meanwhile, for the Type Handling issue, the \emph{review board} agrees with the \emph{code author's} argument that the existing implementation is safe. However, it's concerned about potential future changes that could introduce vulnerabilities, as highlighted by the \emph{security researcher agent}. As a result, this issue is classified as partially valid, which is considered benign since it falls below the 
threshold. This example illustrates how the agents collectively separate confirmed vulnerabilities from exaggerated ones, preventing over-reporting and guiding appropriate remediation.

Another example of running \tool{} on vulnerable code can be found in Appendix H~\cite{ourreplicationpackageall}. The \emph{security researcher agent} flags the lack of strict checks; the \emph{code author} proposes callbacks to mitigate the issue, and the \emph{review board} ultimately rules it a high-severity, valid vulnerability requiring immediate fix. Here, both the \emph{security researcher} and \emph{code author} align, reinforcing the confidence of the final judgment. This demonstrates \tool{}’s capacity to not only support adversarial debate but also consensus when warranted.

Conversely, in the example of the benign function that can be found in Appendix I~\cite{ourreplicationpackageall}, the \emph{security researcher agent} initially raises a concern about certificate validation. The \emph{code author agent} points out that X.509 verification is already used, which the review board agent independently confirms. The issue is ruled invalid. This case highlights \tool{}’s strength in filtering false positives as the security researcher agent's overstatement is corrected by technical context from the code author’s agent.

In another example of benign function (Appendix J~\cite{ourreplicationpackageall}), the \emph{security researcher agent} flags potential leakage via verbose logging. The \emph{code author} proposes mitigations: increasing verbosity thresholds, securing log storage, and reviewing logged content. The \emph{moderator} captures this exchange. After reviewing the evidence, the \emph{review board} classifies the issue as partially valid with low severity (which is below the vulnerability threshold), opting to monitor rather than patch, thereby filtering out an overcautious report. 

Our qualitative analysis shows that the four agents contribute different kinds of knowledge: pattern detection (\emph{security researcher}), code-context refutation or mitigation (\emph{code author}), factual summarisation (\emph{moderator}), and balanced final judgment (\emph{review board}). This minimizes errors on both ends:  they surface genuine threats while filtering out already-handled or overstated concerns, reducing false negatives and false positives.

\begin{tcolorbox}[colback=blue!5!white,colframe=blue!75!black,
         left=2pt,right=2pt,top=2pt,bottom=2pt]
\textbf{Answer to RQ1:} 
\tool{} achieves the highest number of correct function pair predictions while maintaining a balanced precision and recall score. It more than doubles the number of fully‑correct function pairs relative to baselines, while avoiding the extreme false‑positive rates and the low recall. 
\end{tcolorbox}

\subsection{RQ2: Instruction Tuning on Agents}\label{sec:rq2_result}
From the RQ1's results, we observed that \tool{} using GPT-4o achieves the best performance. 
In this RQ, we investigate whether instruction tuning on role-specific agents can further improve \tool{}’s overall performance. 
The results are shown in Table~\ref{tab:rq2_finetuning}.

\paragraph{\textbf{GPT-4o Instruction Tuning.}}
All the single instruction-tuned variants of GPT-4o in \tool{}, covering the \emph{security researcher agent}, \emph{code author agent}, \emph{moderator agent}, and \emph{review board agent}, outperformed both the single-agent and multi-agent baselines. 
Among these, tuning the \emph{moderator agent} delivered the most substantial performance gain. Specifically, the moderator-tuned version increases the performance of the non-tuned version in \tool{} by 15 P-C scores. Compared to the single-agent baseline, this instruction-tuned version of \tool{} increases P-C score by 56. While for the multi-agent baseline, the improvement of the instruction-tuned version of \tool{} increases P-C score by 52.

\begin{table}[t]
\caption{Results of \tool{} using GPT-4o with instruction-tuned agents: SR (\emph{security researcher}), CA (\emph{code author}), M (\emph{moderator}), and RB (\emph{review board}).}
\label{tab:rq2_finetuning}
\setlength{\tabcolsep}{0.75em}
\begin{tabular}{|l|r|r|r|r|}
\hline
\rowcolor[HTML]{C0C0C0} 
\multicolumn{1}{|c|}{\cellcolor[HTML]{C0C0C0}\textbf{Instruction-tuned}} & \multicolumn{1}{c|}{\cellcolor[HTML]{C0C0C0}\textbf{P-C $\uparrow$}} & \multicolumn{1}{c|}{\cellcolor[HTML]{C0C0C0}\textbf{P-V $\downarrow$}} & \multicolumn{1}{c|}{\cellcolor[HTML]{C0C0C0}\textbf{P-B $\downarrow$}} & \multicolumn{1}{c|}{\cellcolor[HTML]{C0C0C0}\textbf{P-R $\downarrow$}} \\ \hline
None       & 81 &	179 &	125 &	50         \\ \hline
\emph{Security researcher} (SR)         & 54	& 153 &	187	& 41         \\ \hline
\emph{Code author} (CA)         &       52 &	39 &	314	& 30         \\ \hline
\emph{Moderator} (M)         &   \textbf{96} & 	154	& 134	& 51 \\ \hline
\emph{Review board} (RB)         &          72	& 133	& 183	& 47        \\ \hline
SR + CA    & 34	& 40	& 331 &	30      \\ \hline
SR + CA + M         & 32	& 37	& 338	& 28        \\ \hline
SR + CA + M + RB       & 45	& 46	& 305	& 39       \\ \hline
\end{tabular}
\vspace{-0.3cm}
\end{table}

\paragraph{\textbf{Effect of Agent-Specific Tuning.}} 
Upon further analysis of the instruction-tuned \emph{moderator agent}, we observed that its summaries are more concise compared to the non-tuned version. 
Specifically, the tuned version of the moderator agent provides a shorter summary than the non-tuned version, reducing the summary length for \emph{security researchers} by 16.03\% and for \emph{code authors} by 16.58\%.
In cases where the tuned \tool{} correctly classified code as benign but the non-tuned version did not, the \emph{security researcher’s} summary was 21.31\% shorter in the tuned version.  Meanwhile, when the tuned \tool{} correctly classified code as vulnerable but the non-tuned version did not, the \emph{code author’s} summary was 19.03\% shorter. 
These observations suggest that the instruction-tuned \emph{moderator agent} can effectively distill key points from both the \emph{security researcher} and \emph{code author}, which help the \emph{review board agent} to make more accurate final verdicts.

We next examined the impact of instruction tuning on the \emph{security researcher} and \emph{code author} agents. Although both configurations still surpassed single-agent and prior multi-agent baselines, their performance fell below that of the non-tuned \tool{}. 
After tuning, we found that the \emph{security researcher} agent proposed fewer potential vulnerabilities, with an average of 2.58 per code instance compared to 4.11. 
To generate the tuning data, GPT-4o is provided with ground truth information, such as CVE details. Since it already knows the ground truth vulnerabilities, the number of potentially vulnerable instances identified by \emph{security researcher agent} is smaller. On average, \emph{security researcher} identify 1.25 vulnerabilities in the training data, which is lower compared to the non-tuned version.  
This low number of vulnerabilities in the tuning data leads to a lower number of vulnerabilities identified in the tuned version of \emph{security researcher agent}.

Similarly, instruction tuning the \emph{code author agent} altered its mitigation-to-refutation ratio. 
While the mitigation to refutation ratio of the non-tuned version was  2:1, the tuned version exhibited a 1:4 ratio. 
In other words, 
the tuned \emph{code author} refuted vulnerabilities more frequently.
From its tuning data, 
it learned to doubt potential vulnerabilities highlighted by \emph{security researcher}. 
Although this approach can restrict over-diagnosing vulnerabilities, it risks undervaluing genuine security concerns, as highlighted by an increase in P-B.  
Despite outperforming the single-agent baseline, this shows the delicate balance required for instruction tuning.  While instruction tuning can sharpen an individual agent’s skills, it may introduce biases or extreme behaviors that undermine collective performance.

\begin{table}[t]
\begin{threeparttable}
\caption{Results of \tool{} using GPT-4o 
across different numbers of discussion rounds. 
}
\label{tab:rq3_rounds}
\begin{tabular}{|l|l|r|r|r|r|}
\hline
\rowcolor[HTML]{C0C0C0} 
\multicolumn{1}{|c|}{\cellcolor[HTML]{C0C0C0}\textbf{Model}}        & \multicolumn{1}{c|}{\cellcolor[HTML]{C0C0C0}\textbf{\#Round}} & \multicolumn{1}{c|}{\cellcolor[HTML]{C0C0C0}\textbf{P-C $\uparrow$}} & \multicolumn{1}{c|}{\cellcolor[HTML]{C0C0C0}\textbf{P-V $\downarrow$}} & \multicolumn{1}{c|}{\cellcolor[HTML]{C0C0C0}\textbf{P-B $\downarrow$}} & \multicolumn{1}{c|}{\cellcolor[HTML]{C0C0C0}\textbf{P-R $\downarrow$}} \\ \hline
     & 1       & \textbf{96} & 	154	& 134 & 51        \\ \cline{2-6} 
     & 2       & 87	& 173	& 122	& 53        \\ \cline{2-6} 
\multirow{-3}{*}{GPT-4o$^\ddagger$} & 3       & 81	& 194	& 110 &	50        \\ \hline
\end{tabular}
\begin{tablenotes}
  \small
  \item $\ddagger$ Instruction tuning for \emph{moderator agent}. 
\end{tablenotes}
\vspace{-0.75em}
\end{threeparttable}
\vspace{-0.5em}
\end{table}

\paragraph{\textbf{Combining Agent-Specific Tuning.}} We also explored configurations with multiple tuned agents, such as the \emph{security researcher} and \emph{code author}, are used together in \tool{}. 
These combinations did not surpass the performance achieved by tuning only the \emph{moderator agent}. 
For instance, when all agents were instruction-tuned, the average number of identified vulnerabilities dropped from 4.11 to 2.60, and the refutation-to-mitigation ratio shifted to 1:4. This imbalance corresponded to a higher P-B (331). 
These lower results may stem from the instruction tuning setup, where the review board is trained on histories that include inputs from other agents.  In the tuning data, the review board often learns to ``follow'' the code author's position (especially strong arguments), which is aligned with the ground truth.  Once the code author is instruction-tuned and becomes more assertive in refuting vulnerabilities, the review board (also fine-tuned) tends to side with the author. Future work should decouple agent training, for example, including cases where the review board correctly identifies a vulnerability despite strong refutations, to reduce over-alignment and improve decisions.

\begin{tcolorbox}[colback=blue!5!white,colframe=blue!75!black,left=2pt,right=2pt,top=2pt,bottom=2pt]
\textbf{Answer to RQ2:} Instruction tuning for \emph{moderator agent} using 50 paired samples yields the most significant improvements. It improves P-C by 15, 56, and 52 compared to the non-tuned version, single-agent, and multi-agent baselines, respectively.

\end{tcolorbox}

\subsection{RQ3: Effect of Multi Rounds  Discussion}\label{sec:rq3_result}
In the previous RQs, \tool{} used a single round of interaction among the agents. In this RQ, we investigate whether additional discussion rounds enhance performance. 
Table~\ref{tab:rq3_rounds} shows the results of these multiple rounds experiments, showing that the best result is achieved using a single round interaction. 
For GPT-4o, increasing the number of rounds actually led to a decrease in overall performance. Specifically, P-V increases with more rounds. 
From further analysis, we found that the \emph{security researcher agent} suggested fewer potential vulnerabilities in later rounds, with an average of 4.22 per code in Round 1, decreasing to 3.55 in Round 2 and 3.43 in Round 3. 
This decline is also driven by the \emph{code author agent}’s refutations. In 30.88\% of the total predictions, the \emph{security researcher} removed a previously identified vulnerability after the code author disputed it. 
Meanwhile, the \emph{code author agent} became progressively more inclined to acknowledge vulnerabilities, with its mitigation-to-refutation ratio shifting from 1.73:1 in Round 1, to 2.5:1 in Round 2, and finally 3:1 in Round 3. Although the \emph{review board agent} did not blindly accept arguments from either side (for instance, during the three-round setup, in 100 out of 868 code instances, \emph{review board agent} overruled the \emph{code author’s} refutation and decided the vulnerability to be valid), the overall imbalance introduced by these interactions still raised P-V and lower P-C rates.

\begin{tcolorbox}[colback=blue!5!white,colframe=blue!75!black,left=2pt,right=2pt,top=2pt,bottom=2pt]
\textbf{Answer to RQ3:} Using GPT-4o in \tool{} outperforms both single-agent and multi-agent baselines across different rounds of agent interaction (i.e., 1, 2, and 3). However, performance declines when extending beyond a single discussion round.
\end{tcolorbox}

\subsection{RQ4: Ablation Study}\label{sec:rq4_result}
In this RQ, we investigate the necessity of each agent in \tool{}. Table~\ref{tab:rq4} presents the ablation results under different agent-removal scenarios. Overall, we find that all four agents are essential for achieving the best performance.
Removing the \emph{moderator agent} from \tool{} reduces P-C from 96 to 32. This result highlights the moderator’s critical role in summarizing arguments to help \emph{review board agent} in making the decision. This finding is also consistent with the results of RQ2 (see Section~\ref{sec:rq2_result}), where instruction tuning the moderator yielded the largest performance gain. Removing both the \emph{moderator} and \emph{review board agent} causes P-C to further drop to \textcolor{blue}{9,} while P-V (pairs both labeled vulnerable) increases to 407. Highlighting that most of the codes are predicted as vulnerable 
by the \emph{code author}. This imbalance occurs because, without an additional agent \emph{review board}, the final judgment defaults heavily toward labeling codes as vulnerable based on partial agreements from the \emph{code author agent}. Thus, the review board is essential for ensuring balanced final verdicts.

Removing the \emph{code author agent} from \tool{} reduces P-C to 63 and increases P-V to 277. Without the \emph{code author's} counterarguments, the \emph{review board agent} has a limited perspective and frequently accepts vulnerabilities identified by the \emph{security researcher agent}. Removing both the \emph{moderator} and \emph{code author} produces similar results (P-C=63, P-V=269). This underscores the crucial role of the \emph{code author} in ensuring balanced judgments while also demonstrating that the \emph{moderator agent} becomes redundant without opposing arguments to mediate.

\begin{tcolorbox}[colback=blue!5!white,colframe=blue!75!black,left=2pt,right=2pt,top=2pt,bottom=2pt]
\textbf{Answer to RQ4:} Our findings confirm that every agent in \tool{} is necessary to achieve the best performance. 
\end{tcolorbox}

\begin{table}[t]
\caption{Results of ablation study on agent in \tool{} using GPT-4o with moderator agent tuned.}
\label{tab:rq4}
\begin{tabular}{|l|r|r|r|r|}
\hline
\rowcolor[HTML]{C0C0C0} 
\multicolumn{1}{|c|}{\cellcolor[HTML]{C0C0C0}\textbf{Agents}} & \multicolumn{1}{c|}{\cellcolor[HTML]{C0C0C0}\textbf{P-C $\uparrow$}} & \multicolumn{1}{c|}{\cellcolor[HTML]{C0C0C0}\textbf{P-V $\downarrow$}} & \multicolumn{1}{c|}{\cellcolor[HTML]{C0C0C0}\textbf{P-B $\downarrow$}} & \multicolumn{1}{c|}{\cellcolor[HTML]{C0C0C0}\textbf{P-R $\downarrow$}} \\ \hline
Base       & \textbf{96} & 	154	& 134	& 51        \\ \hline
No \emph{moderator}        & 32	& 37	& 338	& 28        \\ \hline
No \emph{code author }         &       63 &	277	& 52 &	43         \\ \hline
No \emph{moderator}+\emph{code author}         &   63	& 269	& 63	& 40  \\ \hline
No \emph{moderator}+\emph{review boards}         &          9	& 407	& 5	& 14        \\ \hline
\end{tabular}
\vspace{-0.3cm}
\end{table}

\section{Discussion}\label{sec:discussion}
\paragraph{\textbf{Cost vs. Performance Trade-offs}}

By assigning specialized roles to agents, mirroring real-world scenarios, \tool{} can facilitate deeper discussions by incorporating diverse rationales from different parties.
However, this multi-agent approach inherently incurs additional costs due to the increased number of input and output tokens.  
Interestingly, while GPT-4o is generally more expensive and performs better than GPT-3.5 as a single-agent, using \tool{} with GPT-3.5 outperforms GPT-4o’s single-agent P-C by 28. This approach also reduces costs, requiring only \$7.46 compared to GPT-4o’s single-agent cost of \$8.10 for the whole testing data. 
Conversely, applying \tool{} to more expensive models, such as GPT-4o, significantly increases costs (approximately \$23.9=32-8.1).  
Therefore, careful consideration is important when selecting models to optimize both cost efficiency and detection performance.

\paragraph{\textbf{Enhancing Role-Specific Agents}}
Tuning the \emph{moderator agent} in \tool{} using GPT-4o improves the baselines and non-tuned version of \tool{}.
These results highlight the potential for improving performance by tuning role-specific agents.
Our study used only a sample of 50 pairs for instruction tuning, keeping the cost at approximately \$17 for each agent. 
This is significantly lower than using the instruction tuning setting done in PrimeVul~\cite{ding2024vulnerability}, which would cost more than \$500 for GPT-4o (>15,000 samples). 

\paragraph{\textbf{Analysis on Instruction Tuning Data.}}
Comparing the 50 fine-tuning pair samples from the validation dataset with the 96 successful \tool{} pairs, we found that 39 projects from the successful cases do not appear in the fine-tuning set. 
Only 8 projects are shared. 
Of these, 3 projects contain only commits that occurred after the corresponding training commits. 
In total, only 11 successful pairs involve commits made earlier than the latest commit used in training. 
The fine-tuning data includes 24 unique CWEs, 20 of which also appear in the successful set.  
12 CWEs in the successful set are not covered by the fine-tuning data. These results show minimal overlap in terms of projects, commit timing, and CWE types, suggesting that the model’s performance is not a result of memorization.

\begin{table}[t]
\caption{Results of running \tool{} in the wild.}
\label{tab:vultrial_wild}
\resizebox{0.48\textwidth}{!}{%
\begin{tabular}{|l|r|r|r|r|}
\hline
\rowcolor[HTML]{C0C0C0}
\textbf{Project} & 
\shortstack[c]{\textbf{Funcs}\\\textbf{analyzed}} & 
\shortstack[c]{\textbf{Alarms}\\\textbf{raised}} & 
\shortstack[c]{\textbf{False}\\\textbf{positives}} & 
\shortstack[c]{\textbf{True}\\\textbf{positives}} \\ \hline
Superset~\cite{superset}           & 47  & 0 & 0 & 0 \\ \hline
DataHub~\cite{datahub}   & 160 & 2 & 0 & 2 \\ \hline
InvenTree~\cite{inventree}      & 11  & 1 & 0 & 1 \\ \hline
\end{tabular}
}
\vspace{-0.2cm}
\end{table}

\paragraph{\textbf{\tool{} in the Wild}}
To assess \tool{} in real-world settings, we conducted a pilot experiment on selected open-source projects. Specifically, we targeted well-maintained repositories with active security practices. To identify such projects, we filtered GitHub repositories that include a SECURITY.md file and have over 1,000 stars. We chose the projects that have a SECURITY.md file, as it typically indicates that maintainers are attentive to security concerns and have set up a process for reporting vulnerabilities.

This selection process yielded 196 repositories. From this set, we randomly chose three projects for analysis. For each project, we analyzed recent commits, extracted modified functions, and applied \tool{} with its default rule set, no project-specific tuning. The results of this experiment are summarized in Table~\ref{tab:vultrial_wild}. We manually inspected the flagged vulnerabilities, attempted exploit generation, and reported confirmed issues to the respective maintainers if the vulnerable code was used in the project (i.e., not dead code).

In InvenTree, we discovered a denial-of-service vulnerability in the built-in \texttt{label\_sheet}  plugin. The plugin accepts a \texttt{skip} parameter with no upper bound, allowing an attacker to allocate excessively large Python lists and exhaust server memory. This issue affects any authenticated user who can request labels and has been assigned CVE-2025-49000.
In DataHub, \tool{} identified two functions with resource-exhaustion risks (CWE-400). One function is still actively used, while the other is no longer in use but remains in the codebase. We disclosed the first issues following instructions on their SECURITY.md, and the DataHub developers confirmed the findings.
These results demonstrate that \tool{} is capable of detecting previously unknown vulnerabilities in the wild. To reduce the burden on maintainers, we adopted a conservative reporting strategy and only reported issues we could confidently confirm as true vulnerabilities. We have made the results, vulnerability reports, and exploits available~\cite{ourreplicationpackage}.

\paragraph{\textbf{Generalizability of \tool{}}}
In our main experiments, \tool{}, using OpenAI's models, outperforms the baseline methods.  
Next, to investigate its generalizability, we further evaluated it with a different model family, specifically, LLaMA-3.1-8B~\cite{dubey2024llama}. 
Table~\ref{tab:results-llama} shows the results, indicating the consistent performance of \tool{} even with a different model family. 
CoT struggles under LLaMA-3.1-8B, with only 4 correct predictions (P-C) and a recall of 0.02, 
largely due to a high number of benign misclassifications (P-B = 425). 
\tool{} improves recall to 0.48 and raises P-C to 89, although it has an increased false-positive rate (FPR = 0.45). This suggests that \tool{} favors broader vulnerability coverage, potentially at the cost of mislabeling some benign pairs.

When compared with GPTLens, \tool{} presents a different trade-off. GPTLens achieves higher recall (0.55) and fewer benign misclassifications (P-B = 103), but it also produces more P-V (158) and a higher FPR (0.58). \tool{}, in contrast, has more correct predictions overall (P-C = 89 vs. 80) and higher precision (0.52 vs. 0.49), while maintaining a lower false-positive rate (0.45). These differences indicate that \tool{} may provide a more conservative alternative to GPTLens, with improved balance across evaluation metrics. 
Overall, these findings suggest that \tool{}'s multi-agent reasoning approach can generalize to other models.

\begin{table}[t]
\caption{Results of \tool{}, single-agent, and multi-agent baselines with LLaMA-3.1-8B as the base model.}
\label{tab:results-llama}
\resizebox{0.5\textwidth}{!}{%
\begin{tabular}{|l|r|r|r|r|r|r|r|r|}
\hline
\rowcolor[HTML]{C0C0C0} 
\multicolumn{1}{|c|}{\cellcolor[HTML]{C0C0C0}\textbf{Method}} & \multicolumn{1}{c|}{\cellcolor[HTML]{C0C0C0}\textbf{P-C$\uparrow$}} & \multicolumn{1}{c|}{\cellcolor[HTML]{C0C0C0}\textbf{P-V$\downarrow$}} & \multicolumn{1}{c|}{\cellcolor[HTML]{C0C0C0}\textbf{P-B$\downarrow$}} & \multicolumn{1}{c|}{\cellcolor[HTML]{C0C0C0}\textbf{P-R$\downarrow$}} & \multicolumn{1}{c|}{\cellcolor[HTML]{C0C0C0}\textbf{E$\downarrow$}} & \multicolumn{1}{c|}{\cellcolor[HTML]{C0C0C0}\textbf{P$\uparrow$}}& \multicolumn{1}{c|}{\cellcolor[HTML]{C0C0C0}\textbf{R$\uparrow$}}& \multicolumn{1}{c|}{\cellcolor[HTML]{C0C0C0}\textbf{FPR$\downarrow$}}\\ \hline
CoT\cite{ding2024vulnerability} & 4 & 3 & 425 & 3 & 431 & 0.54 & 0.02 & 0.01 \\ \hline
GPTLens\cite{gptlens} & 80 & 158 & 103 & 94 & 355 & 0.49 & 0.55 & 0.58 \\ \hline
\tool{} & 89 & 120 & 150 & 76 & 346 & 0.52 & 0.48 & 0.45 \\ \hline
\end{tabular}
}
\vspace{-0.2cm}
\end{table}

\paragraph{\textbf{Evaluating Explanation}}
\tool{} provides explanations for each agent’s decisions, helping developers understand why a function is classified as vulnerable or benign rather than receiving a simple binary outcome. Because multiple agents contribute, developers may need to read the full multi-agent discussion to grasp the reasoning. 
To improve readability, we use GPT-4o to synthesize all agent reasoning into a single coherent explanation. We compare the synthesized \tool{} explanations with CoT-based ones using four metrics (i.e., completeness, informativeness, clarity, and actionability), each scored on a 1-5 scale~\cite{widyasari2025explaining,degen2022explain, zhou2021evaluating,velmurugan2025developing}. 
For fairness, we evaluate only cases where the P-C intersections of \tool{} and CoT match under GPT-4o. We conduct automated evaluation using a closed-source LLM from a different family (Claude-Haiku-4.5~\cite{anthropic2025claudeHaiku}) and manual evaluation by three external annotators on 20 instances, without revealing explanation sources. Both evaluations show that synthesized \tool{} explanations outperform CoT across all metrics (Table~\ref{tab:eval_exp}). All generated explanations and evaluation results are available in the replication package~\cite{ourreplicationpackageall}.

\section{Threats to Validity}\label{sec:threats}
\noindent \textbf{\textit{External Validity.}} 
In this study, we focused the evaluation using PrimeVul pair data, which is a challenging recently-released vulnerability detection task on C/C++. 
We did not use other datasets as they may not be free from bias that have been uncovered by Ding et al.~\cite{ding2024vulnerability}. 
In the future, we plan to investigate \tool{} performance using other clean datasets. Another threat is the risk of data leakage, as ChatGPT’s training data is not publicly disclosed.  We mitigate this by evaluating \tool{} using LLaMA-3.1-8B, whose training cutoff (Dec 2023) is before PrimeVul’s GitHub release (Mar 2024). \tool{} also identifies previously unknown real-world vulnerabilities, including CVE-2025-49000, showing it can generalize beyond training data. Therefore, we believe the threats are minimal.

\noindent \textbf{\textit{Construct Validity.}}  
In this study, we use both pair-wise metrics and traditional metrics in our experiments (Section~\ref{sec:metrics}). 
These metrics provide a reliable measure of comparative performance, minimizing the threat to construct validity.
In our experiments,   \tool{} uses a single underlying model for all agents (e.g., GPT-4o, and LLaMA-3.1-8B), we rely on prompt-based role specialization to induce diverse behaviors (e.g., prosecutor vs. defense attorney). Prior works~\cite{wei2022chain,chatterjee2024posix,hu2023large,shinn2023reflexion,wu2024autogen} have shown that such prompt-induced roles can lead to qualitatively different reasoning paths even within the same model. In future work, we plan to use different LLMs for each agent role to further enhance diversity.

\noindent \textbf{\textit{Internal Validity.}} Threats to internal validity in our study refer to possible errors in our experiment. 
To mitigate the risk, we have carefully checked the experiments that we run in our study. The code, prompt, and results are also included in our replication package~\cite{ourreplicationpackageall}.
Furthermore, we set the temperature to 0 to reduce the randomness effect of the experiment. 
We also did the preliminary evaluation where we ran the \tool{} on testing data three times and found that the standard deviation is low (0.7).

\begin{table}[t]
\caption{Results of automatic and manual evaluations of the generated explanation from CoT and synthesized \tool{}.}
\label{tab:eval_exp}
\resizebox{0.48\textwidth}{!}{%
\begin{tabular}{l|cccc|}
\cline{2-5}
                             & \multicolumn{4}{c|}{\cellcolor[HTML]{C0C0C0}\textbf{CoT / Synthesized \tool{}}}                                                                                                                                                                            \\ \cline{2-5} 
                             & \multicolumn{1}{c|}{\cellcolor[HTML]{C0C0C0}\textbf{Completeness}} & \multicolumn{1}{c|}{\cellcolor[HTML]{C0C0C0}\textbf{Clarity}} & \multicolumn{1}{c|}{\cellcolor[HTML]{C0C0C0}\textbf{Actionability}} & \cellcolor[HTML]{C0C0C0}\textbf{Inform.} \\ \hline
\multicolumn{1}{|l|}{Claude~\cite{anthropic2025claudeHaiku}} & \multicolumn{1}{c|}{3.15 / 4.10}                                    & \multicolumn{1}{c|}{3.60 / 3.85}                               & \multicolumn{1}{c|}{2.15 / 4.00}                                       & 2.80 / 3.95                                       \\ \hline
\multicolumn{1}{|l|}{Human}   & \multicolumn{1}{c|}{3.83 / 4.32}                                   & \multicolumn{1}{c|}{3.72 / 4.20}                              & \multicolumn{1}{c|}{2.37 / 4.52}                                    & 3.12 / 3.63                                      \\ \hline

\end{tabular}
}
\vspace{-0.4cm}
\end{table}

\section{Related Work}
\label{sec:related_work}
\noindent \textbf{\textit{Single-Agent Vulnerability Detection.}}
Recent work has explored both DL-based and LLM-based methods for vulnerability detection. DL approaches, such as LineVul~\cite{fu2022linevul}, achieve high recall with the use of transformer-based line-level classification. 
Ni et al.~\cite{ni2023distinguishing} proposed a semantic representation learning method, while Steenhoek et al.~\cite{steenhoek2024dataflow} incorporated dataflow. 
In contrast, LLM-based methods have gained traction for their strong results~\cite{gptlens, zhou2024large_emerging}. 
Other studies have explored prompting and CoT ~\cite{tamberg2024harnessing, ullah2023llms, zhou2024large_emerging, nong2024chain, zhang2024prompt, ding2024vulnerability, steenhoek2024comprehensive}.
Tamberg and Bahsi~\cite{tamberg2024harnessing} benchmarked LLMs using advanced prompting techniques, including CoT, tree-of-thought, and self-consistency, finding that CoT with GPT-4 achieved the best performance and outperformed static analysis tools.
Similarly, Zhou et al.~\cite{zhou2024large_emerging} demonstrated that GPT-4 with prompting outperforms fine-tuned CodeBERT. 
Ullah et al.~\cite{ullah2023llms} evaluated commercial LLMs across various prompting strategies, including CoT, and found promising detection performance.
Recent studies have further highlighted the limitations of CoT~\cite{nong2024chain, zhang2024prompt, steenhoek2024comprehensive}. 
Notably, PrimeVul's seminal work found that GPT-3.5 and GPT-4 perform poorly on PrimeVul's pair test set, reinforcing concerns about CoT’s reliability in vulnerability detection. 
Efforts to enhance LLM-based methods include finetuning and retrieval-augmented generation (RAG). 
Shestov et al.~\cite{shestov2025finetuning} showed that finetuning boosts recall but is costly and dataset-sensitive. 
Du et al.~\cite{du2024vul} proposed Vul-RAG, which improves factual grounding via structured knowledge injection. 
Sun et al.~\cite{sun2024llm4vuln} introduced LLM4Vuln, a framework that evaluates and decouples LLM reasoning capabilities. 
In this work, we use DL-based methods and CoT as baselines and show that \tool{} outperforms them on PrimeVul.

\vspace{.25em}

\noindent \textbf{\textit{Multi-Agent Vulnerability Detection.}}
Multi-agent frameworks can enhance LLM-based vulnerability detection by enabling collaboration and specialization among agents~\cite{he2024llm}.
EvalSVA~\cite{wen2024evalsva} introduces a multi-agent framework where agents act as security experts. 
Similarly, MuCoLD~\cite{mao2024multi} agents assume roles such as tester and developer, which collaboratively analyze code. In contrast, \tool{} employs a structured multi-agent design inspired by a courtroom, where agents take on distinct roles such as prosecutor, defender, and judge. This encourages adversarial debate and balanced reasoning that leads to robust decision-making.
We excluded these recent works as baselines due to the lack of replication packages and insufficient prompt details to replicate the study. 

Other multi-agent-based approaches have been explored for smart contract auditing~\cite{wei2024llm, ma2024combining, hu2023large}.
LLM-SmartAudit~\cite{wei2024llm} uses collaborative decision-making with role-play and reasoning to detect vulnerabilities in smart contracts.
Similarly, iAudit~\cite{ma2024combining} assigns specific roles, such as detector, reasoner, and critic, to guide the analysis of Solidity smart contracts.
GPTLens~\cite{hu2023large} employs multiple LLM-based agents acting as independent prosecutors to identify vulnerabilities, with a critic agent reviewing and ranking the findings.
Our work differs from these studies in two key ways.
First, we focus on vulnerability detection in C and C++ on PrimeVul, a benchmark comprising challenging real-world examples.
Second, \tool{} adopts a structured, courtroom-inspired framework specifically tailored for general-purpose languages beyond smart contracts.
As these studies focus on smart contracts, their scope differs from our work for general-purpose programming languages.
Nonetheless, we found that GPTLens can be adapted to general-purpose vulnerability detection and included it as a baseline in our experiments (Section~\ref{sec:model}).
Our results demonstrate that \tool{} substantially outperforms GPTLens and single-agent baselines on PrimeVul.

\section{Conclusion and Future Work}\label{sec:conclusion}
We proposed \tool{}, a novel explainable multi-agent LLM-based approach for vulnerability detection that applies a courtroom-style paradigm. \tool{} comprises four role-specific agents: the \emph{security researcher} (prosecutor), \emph{code author} (defense attorney), \emph{moderator} (judge), and \emph{review board} (jury). Each agent is essential to get the best performance (Section~\ref{sec:rq4_result}). We evaluated \tool{} against single-agent (CodeBERT~\cite{feng2020codebert}, CodeT5~\cite{wang2021codet5}, UniXCoder~\cite{guo2022unixcoder}, LineVul~\cite{fu2022linevul}, and Ding et al.'s CoT~\cite{ding2024vulnerability}) and multi-agent (GPTLens~\cite{hu2023large}) baselines. Our results show that \tool{} outperforms both the single-agent and multi-agent baselines, with instruction tuning in the \emph{moderator agent} further enhancing its performance.
We have also demonstrated the efficacy of \tool{} across different LLMs, including an open-source, in-house-deployable model (LLaMA-3.1-8B), as well as the high quality of its generated explanations and its ability to uncover multiple zero-day vulnerabilities.

For future work, we plan to apply \tool{} to more datasets to evaluate its generalizability further. We also plan to consider the strengths and weaknesses of different LLMs to choose the one that best fits as the base model for each role-specific task. 

\begin{acks}
This research / project is supported by the National Research Foundation, Singapore, and the Smart Nation Group under the Smart Nation Group's Translational R\&D Grant (Award No. TRANS2023-TGC02). Any opinions, findings and conclusions or recommendations expressed in this material are those of the author(s) and do not reflect the views of National Research Foundation, Singapore or the Smart Nation Group.
\end{acks}

\balance
\bibliographystyle{ACM-Reference-Format}
\bibliography{sample-base, software}

\end{document}